\documentclass[12pt]{article} 
\usepackage{graphicx,subfigure,latexsym,amssymb}
\usepackage{bm}

\newcommand{\be}{\begin{equation}}
\newcommand{\ee}{\end{equation}}
\newcommand{\bear}{\begin{eqnarray}}
\newcommand{\eear}{\end{eqnarray}}
\newcommand{\ba}{\begin{array}}
\newcommand{\ea}{\end{array}}

\def\({\left(}
\def\){\right)}

\begin{document}

\begin{titlepage}
\vfill
\begin{flushright}
{\normalsize RBRC-1246}\\
\end{flushright}

\vfill
\begin{center}
{\Large\bf  Dynamic Universality Class of Model H with Frustrated Diffusion: $\epsilon$-Expansion}

\vskip 0.3in
Ho-Ung Yee$^{1,2}$\footnote{e-mail:
{\tt hyee@uic.edu}}
\vskip 0.3in

 {\it $^{1}$Department of Physics, University of Illinois,} \\
{\it Chicago, Illinois 60607 }\\[0.15in]
{\it $^{2}$RIKEN-BNL Research Center, Brookhaven National Laboratory,} \\
{\it Upton, New York 11973-5000 }
\\[0.3in]

{\normalsize  2017}

\end{center}

\vfill

\begin{abstract}

We study a variation of the dynamic universality class of model H in a spatial dimension of $d=4-\epsilon$, by frustrating charge diffusion and momentum density fluctuations along $d_T=1$ or $d_T=2$ dimensions, while keeping the same dynamics of model H in the other $d_L=d-d_T$ dimensions. The case of $d_T=2$ describes the QCD critical point in a background magnetic field. We find that these models belong to a different dynamical universality class due to extended conservation laws compared to the model H, although the static universality class remains the same as the 3-dimensional Ising model. We compute the dynamic critical exponents of these models in first order of $\epsilon$-expansion to find that $x_\lambda\approx 0.847\,\epsilon$, $x_{\bar\eta}\approx 0.153\,\epsilon$, and $z=4-x_\lambda\approx 3.15$ when $\epsilon=1$ and $d_T=2$. 
For $d_T=1$ the results are numerically similar to the model H values: $z\approx 3.08$.

\end{abstract}

\vfill

\end{titlepage}
\setcounter{footnote}{0}

\baselineskip 18pt \pagebreak
\renewcommand{\thepage}{\arabic{page}}
\pagebreak

\section{Introduction \label{sec1}}

In this work we study a particular variation of dynamic universality class of model H \cite{Siggia:1976zz,Hohenberg:1977ym} in the $\epsilon$-expansion \cite{Wilson:1971dc,Wilson:1973jj}. Recall that the model H has two dynamical fields: a conserved charge density and the shear component of momentum density fluctuations. Their dynamics in long wave-lengths near the critical regime is mainly governed by diffusion dynamics due to their conservation laws.
What gives a non-trivial IR fixed point is a non-linear coupling between these two diffusion dynamics \cite{Kadanoff:1968zz,kawasaki}.
This coupling constant at the fixed point is of order $\epsilon$, and a perturbation theory in terms of $\epsilon$-expansion is possible. The results for dynamical critical exponents such as $x_\lambda$, $x_\eta$ and $z$ are available up to $\epsilon^2$ \cite{Siggia:1976zz}. When $\epsilon=1$, they are $x_\lambda\approx 1$, $x_\eta\approx 0$ and $z\approx 3$.

The model we study is a simple twist of the model H: we will frustrate the charge conductivity and the momentum density fluctuations along $d_T=1$ or $d_T=2$ spatial dimensions (denoted as $\bm x_T$), while keeping them in the rest $d_L=4-d_T-\epsilon$ dimensions (denoted as $\bm x_\parallel$). There are at least two motivations to consider these:
1) $d_T=1$ case can describe the layered system where the charge transport and the momentum flow across the layers are frustrated or damped with a finite relaxation time, so that these modes decouple near the critical regime, 2) $d_T=2$ case describes a system in a background magnetic field $F_{12}\neq 0$, where
momentum fluctuations in $(1,2)$ directions are damped with a finite relaxation time and are no longer a hydrodynamic variable \cite{Li:2017tgi}. In strong magnetic field limit, a charge conductivity along $(1,2)$ directions should be suppressed as well due to small cyclotron orbits. The $d_T=2$ case should describe the dynamic universality class for the QCD critical point \cite{Stephanov:1999zu,Berdnikov:1999ph,Son:2004iv} in a strong background magnetic field.

Let us explain the case 2) in some more detail \cite{Li:2017tgi}.  The hydrodynamic equations with a background magnetic field are
\be
{J}^\mu=\lambda F^{\mu\nu}u_\nu\,,\quad \partial_{\mu}T^{\mu\nu}=F^{\nu\alpha}J_\alpha\,,
\ee 
where $J^\mu$ is the charge current, $T^{\mu\nu}=wu^\mu u^\nu+p\eta^{\mu\nu}$ is the energy-momentum, $u^\mu$ is the fluid velocity field and $\lambda$ is the conductivity. For small transverse velocity $v_1, v_2$, the first equation gives the current $J^i=\lambda F_{12}\epsilon^{ij}v_j$ ($i,j=1,2$).
In the homogeneous limit ($\bm k\to 0$), the second equation gives
\be
\partial_t v_i=
-{\lambda (F_{12})^2\over w} v_i\equiv -{1\over \tau_R} v_i\,,\label{relax}
\ee
so that $v_{1,2}$ has a finite relaxation time $\tau_R$ in $\bm k\to 0$ limit, that is, it is not a hydrodynamic mode. Once we remove $v_{1,2}$ from the critical dynamical modes, there is no non-linear coupling that gives renormalization for the transverse conductivity along $(1,2)$ directions: the physical transverse conductivity remains finite in infrared. Note that the conductivity appearing in (\ref{relax}) is the transverse conductivity. The naive scaling dimension of the conductivity is negative when $z<4$ (see (\ref{naive})) and the transverse conductivity therefore becomes irrelevant in the infrared in the renormalization group. On the other hand, the physical longitudinal conductivity will be shown to diverge near the critical point due to the non-linear coupling, and its renormalized value goes to a finite fixed point in the renormalization group (see the discussion in section \ref{sec4}). 

We will find that these variations of model H belong to dynamic universality classes that are different from the original model H, and the
dynamic critical exponents such as $x_\lambda$, $x_\eta$ and $z$ are different: we compute these exponents in first order of the $\epsilon$-expansion by using the Wilsonian renormalization group method \cite{Wilson:1971dc,Wilson:1973jj}. Using the Feynman diagram method of Ref.\cite{Wilson:1971vs} one can in principle continue to higher orders in $\epsilon$.
The reason why we get the different dynamic universality classes in these models is that the symmetries are enlarged from those of the original model H. In the absence of charge diffusion along $d_T$, the integrated charge density along $d_L$ at any given point in $d_T$ is conserved: the symmetry group is infinite dimensional.

Despite that the real-time dynamics in these models breaks rotational symmetry,
we assume that the static universality class remains to be the isotropic 3D Ising model described by a theory of a scalar field $\psi$ with quartic interaction.
This is reasonable since the only relevant or marginal perturbation near critical regime that breaks rotational symmetry is of a form $K^{ij}\partial_i\psi \partial_j\psi$ and we can make it isotropic by diagonalizing and rescaling the coordinates.

\section{Description of the model}

We will follow the notations in Ref.\cite{Siggia:1976zz,Hohenberg:1977ym}, and denote a conserved order parameter by $\psi$ and the momentum density vector by $\bm j_\parallel$ which has components only along the space $\bm x_\parallel$ of dimension $d_L=4-\epsilon-d_T$ in our model. The static equilibrium thermal distribution is
given by $e^{-F[\psi,\bm j_\parallel]}$ where $F$ is the free energy functional of 3D Ising universality class
\be
F=\int d^{4-\epsilon}x \left({1\over 2}(\partial\psi)^2+{r \Lambda^2\over 2}\psi^2+{u\Lambda^{\epsilon}\over 4!}\psi^4 +{1\over 2}\bm j_\parallel\cdot\bm j_\parallel\right)\,,
\ee
where $\Lambda$ is the physical UV-cutoff in momentum space. Writing the parameters of the model in the above way, $(r,u)$ are dimensionless.
The dynamical model is a simple twist of the model H with frustrated diffusion along $\bm x_T$,
\bear
{\partial\psi\over\partial t}&=&\lambda_\parallel\Lambda^{z-4}\nabla_\parallel^2 {\delta F\over\delta \psi}-g\Lambda^{z-3+\epsilon/2}\bm\nabla_\parallel\psi\cdot{\delta F\over\delta\bm j_\parallel}+\theta\,,\\
{\partial\bm j_\parallel\over\partial t}&=& {\cal P}\left(\bar\eta_\perp\Lambda^{z-2}\nabla_\perp^2\bm j_\parallel+\bar\eta_\parallel\Lambda^{z-2}\nabla^2_\parallel\bm j_\parallel+g\Lambda^{z-3+\epsilon/2}\bm\nabla_\parallel\psi {\delta F\over\delta \psi}+\bm\xi\right)\,,
\eear
where ${\cal P}$ projects the vector components onto the subspace perpendicular to a momentum vector $\bm k$ in Fourier space, that is, it keeps only the shear components. Since $\bm j_\parallel$ is already perpendicular to $\bm k_\perp$, it is practically a projection operator in $\bm x_\parallel$ space;
\be
{\cal P}\to \delta^{ij}_\parallel-{\bm k^i_\parallel \bm k^j_\parallel\over \bm k_\parallel^2}\,.
\ee
The random noises $(\theta,\bm\xi)$ are Gaussian with the strength determined by Fluctuation-Dissipation relation,
\bear
\langle \theta(\bm x,t)\theta(\bm x',t')\rangle&=&-2\lambda_\parallel\Lambda^{z-4}\nabla_\parallel^2 \delta(\bm x-\bm x')\delta(t-t')\,,\nonumber\\
\langle \bm\xi^i(\bm x,t)\bm\xi^j(\bm x',t')\rangle&=&-2\delta^{ij}\left(\eta_\perp\Lambda^{z-2}\nabla_\perp^2+\bar\eta_\parallel\Lambda^{z-2}\nabla_\parallel^2\right)\delta(\bm x-\bm x')\delta(t-t')\,,
\eear
which ensures that the equilibrium distribution is $e^{-F}$.
The $\lambda_{\parallel}$ is the conductivity, $\bar\eta_{\parallel,\perp}$ are the shear viscosities in $\bm x_{\parallel}$ and $\bm x_\perp$ spaces respectively, which are in general different. We will see that their values at the IR fixed point are indeed different, but comparable to each other so we need to keep them both. The $g$ is the non-linear coupling between two dynamical modes in model H, which drives a non-trivial IR fixed point. The exponent $z$ is the scale dimension of the frequency $\omega$ relative to the momentum $\bm k$. All parameters defined as above are then dimensionless.

Although we can perform the Wilsonian renormalization group at the level of the above equations of motion, we choose to work in a language of stochastic field theory or a path integral, where the renormalization procedure looks more organized (at least to the eyes of the author).
For that purpose, we introduce the ``a-type'' fields $(\psi_a,\bm j_{\parallel a})$ and call the original variables in the equations of motion the ``r-type'' fields $(\psi_r,\bm j_{\parallel r})$, then consider a path integral of $e^S$ with an action $S=\int dt\int d^{4-\epsilon} x \,\,{\cal L}$,
\bear
{\cal L}&=&i\psi_a\left({\partial\psi_r\over \partial t}-\lambda_\parallel \Lambda^{z-4}\nabla_\parallel^2 {\delta F\over\delta\psi_r}+g\Lambda^{z-3+\epsilon/2}\bm\nabla_\parallel\psi_r\cdot{\delta F\over\delta\bm j_{\parallel r}} -\theta\right)\nonumber\\
&+& i\bm j_{\parallel a}\cdot{\cal P}\left({\partial\bm j_{\parallel r}\over\partial t}-\bar\eta_\perp\Lambda^{z-2}\nabla_\perp^2\bm j_{\parallel r}-\bar\eta_\parallel\Lambda^{z-2}\nabla^2_\parallel\bm j_{\parallel r}-g\Lambda^{z-3+\epsilon/2}\bm\nabla_\parallel\psi_r {\delta F\over\delta \psi_r}-\bm\xi\right)\nonumber\\
&-&{1\over 2}\theta {1\over -2\lambda_\parallel\Lambda^{z-4}\nabla_\parallel^2}\theta-{1\over 2}\bm\xi\cdot {{\cal P}\over -2\left(\eta_\perp\Lambda^{z-2}\nabla_\perp^2+\bar\eta_\parallel\Lambda^{z-2}\nabla_\parallel^2\right)}\bm\xi\,.
\eear
The correspondence to the usual $(r,a)$-fields in the Schwinger-Keldysh formalism is clear: the ``r-type'' fields are classical variables.
The path integral over the ``a-type'' field localizes the path integral of ``r-type'' fields to the solutions of the original stochastic Langevin equations of motion with the Gaussian random noises $(\theta,\bm\xi)$. 
The ``wave-function'' at time $t$ is precisely then the probability distribution of the classical variables at time $t$ generated by solutions of the stochastic equations of motion. 

We can integrate out the noise variables $(\theta,\bm\xi)$ as they are Gaussian to obtain
\bear
{\cal L}'&=&i\psi_a\left({\partial\psi_r\over \partial t}-\lambda_\parallel \Lambda^{z-4}\nabla_\parallel^2 {\delta F\over\delta\psi_r}+g\Lambda^{z-3+\epsilon/2}\bm\nabla_\parallel\psi_r\cdot{\delta F\over\delta\bm j_{\parallel r}} \right)\nonumber\\
&+& i\bm j_{\parallel a}\cdot{\cal P}\left({\partial\bm j_{\parallel r}\over\partial t}-\bar\eta_\perp\Lambda^{z-2}\nabla_\perp^2\bm j_{\parallel r}-\bar\eta_\parallel\Lambda^{z-2}\nabla^2_\parallel\bm j_{\parallel r}-g\Lambda^{z-3+\epsilon/2}\bm\nabla_\parallel\psi_r {\delta F\over\delta \psi_r}\right)\nonumber\\
&+&\lambda_\parallel\Lambda^{z-4}\psi_a\nabla_\parallel^2\psi_a +\bm j_{\parallel a}\cdot\left(\bar\eta_\perp\Lambda^{z-2}\nabla_\perp^2+\bar\eta_\parallel\Lambda^{z-2}\nabla_\parallel^2\right){\cal P}\bm j_{\parallel a}\,.\label{action}
\eear
Upon ``quantization'' the ``a-type'' fields are canonical conjugate to the ``r-type'' fields: $\psi_a\sim i{\partial\over\partial \psi_r}$. The ``Schrodinger equation" is the Fokker-Planck equation. 

What we do in this formulation corresponds to the renormalization in terms of the Fokker-Planck equation, instead of the original stochastic equations of motion. Obviously they should be equivalent.

\section{Renormalization group in $\epsilon$-expansion}

We follow the standard procedure (see Ref.\cite{Siggia:1976zz,Wilson:1971dc,Wilson:1973jj}) of thinning the momentum shell around the cutoff $\Lambda$, and integrate over the shell
\be
\Lambda/b < |\bm k| < \Lambda\,,
\ee 
with a constant $b>1$ close to 1. After this we rescale the coordinates or equivalently the momenta
to get back to the same cutoff $\Lambda$ in the rescaled momentum space $\bm k'$;
\bear
\bm k'&=& b \bm k\,,\nonumber\\
\omega'&=& b^z \omega\,.
\eear 
Without the non-linear couplings such as $g$ and $u$, the parameters of the theory would change at each step simply by their naive scaling dimensions:
\bear
\lambda_\parallel'=b^{z-4}\lambda_\parallel&\approx &\lambda_\parallel+(z-4)\lambda_\parallel\log b\,,\nonumber\\
(\bar\eta_\perp',\bar\eta_\parallel')=b^{z-2}(\bar\eta_\perp,\bar\eta_\parallel)&\approx& (\bar\eta_\perp,\bar\eta_\parallel)+(z-2)(\bar\eta_\perp,\bar\eta_\parallel)\log b\,,\nonumber\\
g'= b^{z-3+\epsilon/2} g&\approx& g+(z-3+\epsilon/2)g\log b\,.\label{naive}
\eear
There are similar equations for the static parameters $(r,u)$.
The non-linear couplings by $(u,g)$ give rise to additional contributions to the above. These are what we need to compute.

In the critical regime, the IR cutoff set by the correlation length $\Lambda_{IR}\sim \xi^{-1}$ is far separated from the UV cutoff $\Lambda$.
After $N$ steps of the above procedure where $b^N\xi^{-1}=\Lambda$, the UV and IR cutoffs in the renormalized theory become comparable and the scaling behavior is lost. This is the point where the hydrodynamic regime sets in, and the renormalization group running of the parameters of the theory stops and further contributions to these parameters are IR-finite. 
The number of steps of the above procedure to be performed to reach this is $N=\log(\Lambda\xi)/\log b$. The renormalized parameter $r$ starts close to its fixed point value which is of order $r^*\sim \epsilon$,
and we can neglect it in the propagator in the leading order perturbation in $\epsilon$. 
In each step of the above procedure, the deviation of the renormalized $r$ from $r^*$ grows, and only after performing the same $N=\log(\Lambda\xi)/\log b$ steps it becomes of order $1$:  this is because the scaling behavior is lost precisely when the renormalized $r$ deviates from $r^*\sim\epsilon$ by order $1$. Therefore $r$ stays of order $\epsilon$ in most of the $N$-steps, and we can neglect $r\ll 1$ in the propagators in all $N$-steps in the leading perturbation in $\epsilon$-expansion \cite{Wilson:1971dc,Wilson:1973jj}.

When $z>1$, the cutoff in frequency space can be taken to be infinite.
Even if we started with a same cutoff $\Lambda$ in the frequency space, a step of above procedure would change it to $b^{z-1}\Lambda$. In the critical regime where $N\gg 1$, this cutoff quickly becomes much larger than $\Lambda$ in most of the $N$ steps.


The observed physical parameters of the theory are the ones without rescaling the coordinates and the time: they are the parameters measured in terms of the original coordinates and time. The effect of rescaling coordinates for them is simply given by (\ref{naive}), and therefore the observed physical parameters are obtained from the renormalized parameters by undoing the naive scaling transformation (\ref{naive}) \cite{Siggia:1976zz,Wilson:1971dc,Wilson:1973jj}:
\bear
\lambda_\parallel^{phys}&=&(b^{4-z})^N \lambda_\parallel^*\,,\nonumber\\
(\bar\eta_\perp^{phys},\bar\eta_\parallel^{phys})&=& (b^{2-z})^N (\bar\eta_\perp^*,\bar\eta_\parallel^*)\,,\nonumber\\
g^{phys}&=&(b^{3-z-\epsilon/2})^N g^*\,,
\eear
where the starred parameters are the renormalized parameters after performing the large $N$-steps (they could be a finite fixed point value or not).
In effect, the physical parameters in the above capture only the contributions from the non-linear couplings in the relevant momentum range $\xi^{-1}<|\bm k|<\Lambda$ as they should.

With these general discussions reviewed, we now compute the contributions from the non-linear couplings $(g,u)$ to the renormalization group equation. We can do a perturbation theory in these couplings since their fixed point values will be of order $\epsilon$. After integrating over the momentum shell, the action $S+\delta S$ with a new cutoff $\Lambda/b$ is expected to be given by a new set of parameters
\be
(\lambda_\parallel,\bar\eta_{\perp,\parallel})\to (\lambda_\parallel+\delta\lambda_\parallel,\bar\eta_{\perp,\parallel}+\delta \bar\eta_{\perp,\parallel})\,,\quad F\to F+\delta F\,.
\ee
The contribution to $g$ will be shown to be absent.
The contributions from the non-linear couplings $(\delta\lambda_\parallel,\delta\bar\eta_{\perp,\parallel},\delta F)$ are of first order in $\log b$ in small $\log b$ limit.
The effect of rescaling (\ref{naive}) to restore the original cutoff $\Lambda$ and the above non-linear contribution are therefore additive to each other to first order in $\log b$.

First, the $\delta F$ should be identical to what one would have in a static renormalization group \cite{Siggia:1976zz}, because the equilibrium distribution after integrating over the momentum shell is $e^{-(F+\delta F)}$ by Fluctuation-Dissipation relation, and this must agree with what one would have in a static renormalization after integrating over the same momentum shell. Although this is guaranteed by the Fluctuation-Dissipation relation in the original unintegrated theory, it is assuring to see this explicitly 
for a few lowest order contributions, which we have checked.
From the known static renormalization group \cite{Wilson:1971dc,Wilson:1973jj}, we have
\be
F+\delta F=\int d^{4-\epsilon}x\,\,\left((1+\eta\log b){1\over 2}(\partial\psi)^2+{1\over 2}(r+\delta r)\Lambda^2\psi^2+{1\over 4!}(u+\delta u)\Lambda^{\epsilon}\psi^4\right)\,,
\ee
where $\eta/2\sim \epsilon^2$ is the anomalous dimension of $\psi$, and the other terms are not of our interest for the renormalization of the transport coefficients ($r+\delta r$ for example is order $\epsilon$ in the scaling regime and will be neglected in the propagator). 
Looking at how $F$ appears in the charge diffusion term in the action (\ref{action}), this wave-function renormalization
contributes to $\delta\lambda_\parallel$ as
\be
\delta\lambda_\parallel=\eta \lambda_\parallel \log b\,.\label{wavefunction}
\ee

\begin{figure}[t]
	\centering
	\includegraphics[width=10cm]{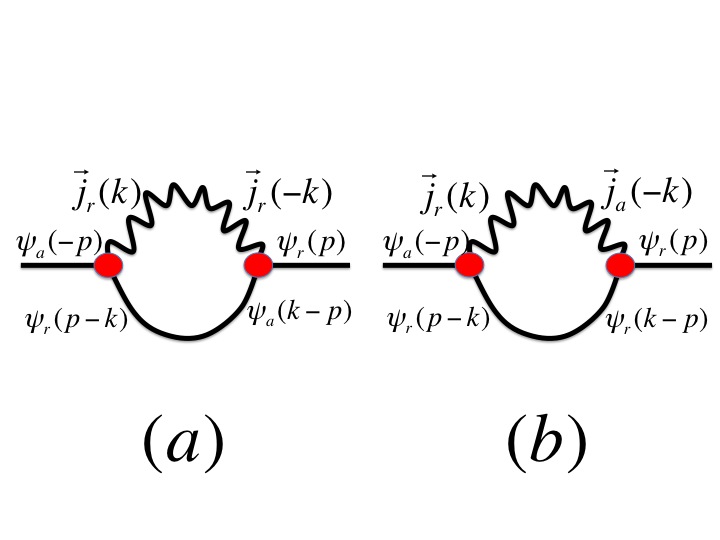}
		\caption{The two real-time diagrams for renormalization of $\lambda_\parallel$. \label{fig1}}
\end{figure}

\begin{figure}[t]
	\centering
	\includegraphics[width=9cm]{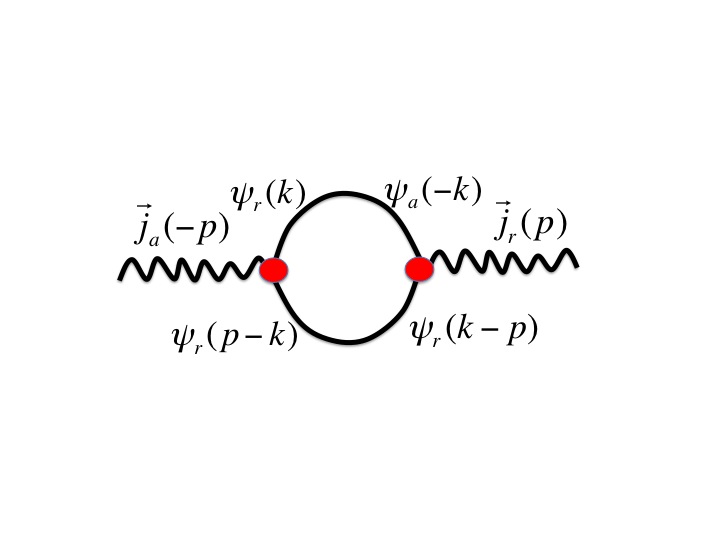}
		\caption{The real-time diagram for renormalization of $\bar\eta_\perp$ and $\bar\eta_\parallel$. \label{fig2}}
\end{figure}

The other leading contributions to $(\delta\lambda_\parallel,\delta\bar\eta_{\perp,\parallel})$ come from
the coupling $g$. One can compute these by looking at the contributions to the $\psi_a\psi_r$ and $\bm j_{\parallel a}\bm j_{\parallel r}$ terms in the action as they contain these transport coefficients.
The real-time Feynman diagrams generating a $\psi_a\psi_r$ term in $\delta S$ are shown in Figure \ref{fig1}.
There are two real-time diagrams to be summed. In Figure \ref{fig2} we show the real-time diagram generating a $\bm j_{\parallel a}\bm j_{\parallel r}$ term in $\delta S$. The solid lines are the propagators of $\psi$-field and the wavy lines are for the $\bm j_\parallel$-fields. These propagators are easily found from the quadratic part of the action $S$ to be
\bear
&&\langle\psi_r(k)\psi_a(-k)\rangle=\langle\psi_a(k)\psi_r(-k)\rangle^*={1\over -\omega-i\lambda_\parallel\Lambda^{z-4}\bm k_\parallel^2\bm k^2}\,,\nonumber\\&&\langle \psi_r(k)\psi_r(-k)\rangle={2\lambda_{\parallel}\Lambda^{z-4}\bm k_\parallel^2\over \omega^2+(\lambda_{\parallel}\Lambda^{z-4}\bm k_{\parallel}^2 \bm k^2)^2}\,,\quad
\langle\psi_a(k)\psi_a(-k)\rangle=0\,,\\
&&\langle\bm j_{\parallel r}^i(k)\bm j_{\parallel a}^j(-k)\rangle=
\langle\bm j_{\parallel a}^i(k)\bm j_{\parallel r}^j(-k)\rangle^*={\left(\delta^{ij}_\parallel-{\bm k_\parallel^i\bm k_\parallel^j\over \bm k_\parallel^2}\right)\over -i\omega-i(\bar\eta_\perp\Lambda^{z-2}\bm k_\perp^2+\bar\eta_\parallel\Lambda^{z-2}\bm k_\parallel^2)}\,,\nonumber\\
&&\langle\bm j_{\parallel r}^i(k)\bm j_{\parallel r}^j(-k)\rangle={2(\bar\eta_{\perp}\Lambda^{z-2}\bm k_\perp^2+\bar\eta_{\parallel}\Lambda^{z-2}\bm k_\parallel^2)\left(\delta^{ij}_\parallel-{\bm k_\parallel^i\bm k_\parallel^j\over \bm k_\parallel^2}\right)\over \omega^2+(\bar\eta_{\perp}\Lambda^{z-2}\bm k_\perp^2+\bar\eta_{\parallel}\Lambda^{z-2}\bm k_\parallel^2)^2}\,,\quad \langle\bm j_{\parallel a}^i(k)\bm j_{\parallel a}^j(-k)\rangle=0\,,\nonumber
\eear
where $k=(\omega,\bm k)=(\omega,\bm k_{\parallel},\bm k_{\perp})$ is a frequency-momentum in Fourier space.
Each vertex in the diagrams is from the $g$ coupling in the action, which is
\be
ig\Lambda^{z-3+\epsilon/2}\psi_a \bm\nabla_\parallel\psi_r\cdot\bm j_{\parallel r}-ig\Lambda^{z-3+\epsilon/2}\bm j_{\parallel a}\cdot\bm\nabla_\parallel\psi_r(-\nabla^2 \psi_r)\,.
\ee

We first write down the expression for the first diagram in Figure \ref{fig1}. 
\bear
&&-g^2\Lambda^{2z-6+\epsilon}\int_p \psi_a(-p)\psi_r(p)\int_k {i(\bm p_\parallel^i-\bm k_\parallel^i) i\bm p_\parallel^j \over -(\Omega-\omega)-i\lambda_\parallel\Lambda^{z-4}(\bm p_\parallel-\bm k_\parallel)^2(\bm p-\bm k)^2}\nonumber\\&\times&{2(\bar\eta_{\perp}\Lambda^{z-2}\bm k_\perp^2+\bar\eta_{\parallel}\Lambda^{z-2}\bm k_\parallel^2)\left(\delta^{ij}_\parallel-{\bm k_\parallel^i\bm k_\parallel^j\over \bm k_\parallel^2}\right)\over \omega^2+(\bar\eta_{\perp}\Lambda^{z-2}\bm k_\perp^2+\bar\eta_{\parallel}\Lambda^{z-2}\bm k_\parallel^2)^2}\,,
\eear
where $p=(\Omega,\bm p)$ is the external momentum, and we denote
\be
\int_p=\int{d\Omega\over 2\pi}\int_{\bm p}=\int{d\Omega\over 2\pi}\int{d^{4-\epsilon}\bm p\over(2\pi)^{4-\epsilon}}\,,
\ee
so that $\int_p \psi_a(-p)\psi_r(p)=\int dt \int d^{4-\epsilon}\bm x\,\psi_a(x)\psi_r(x)$.
Computing $\omega$ integration by closing the contour in the lower half-plane, we obtain
\bear
+ig^2\Lambda^{2z-6+\epsilon}\int_p \psi_a(-p)\psi_r(p){\bm p}_\parallel^2 \int_{\bm k}{\left(1-{(\bm k_\parallel\cdot\bm p_\parallel)^2\over \bm k_\parallel^2\bm p_\parallel^2}\right)\over -i\Omega+(\bar\eta_\perp\Lambda^{z-2}\bm k_\perp^2+\bar\eta_\parallel\Lambda^{z-2}\bm k_\parallel^2)+\lambda_\parallel\Lambda^{z-4}(\bm k_\parallel-\bm p_\parallel)^2(\bm k-\bm p)^2}\,,\nonumber\\
\eear
where $\bm k$ integration to be performed in the momentum shell $\Lambda/b<|\bm k|<\Lambda$.
The second diagram in Figure \ref{fig1} is similarly computed to be
\bear
&&g^2\Lambda^{2z-6+\epsilon}\int_p \psi_a(-p)\psi_r(p)\int_k {i(\bm p_\parallel^i-\bm k_\parallel^i) i(\bm p_\parallel^j (\bm k-\bm p)^2 + (\bm k_\parallel^j-\bm p^j_\parallel)\bm p^2) \over (\omega-\Omega)^2+(\lambda_\parallel\Lambda^{z-4}(\bm p_\parallel-\bm k_\parallel)^2(\bm p-\bm k)^2)^2}\nonumber\\&\times&{2\lambda_\parallel\Lambda^{z-4}(\bm k_\parallel-\bm p_\parallel)^2\left(\delta^{ij}_\parallel-{\bm k_\parallel^i\bm k_\parallel^j\over \bm k_\parallel^2}\right)\over -\omega-i(\bar\eta_{\perp}\Lambda^{z-2}\bm k_\perp^2+\bar\eta_{\parallel}\Lambda^{z-2}\bm k_\parallel^2)}\nonumber\\
&=&-ig^2\Lambda^{2z-6+\epsilon}\int_p \psi_a(-p)\psi_r(p){\bm p}_\parallel^2 \int_{\bm k}{\left(1-{(\bm k_\parallel\cdot\bm p_\parallel)^2\over \bm k_\parallel^2\bm p_\parallel^2}\right)\left(1-{\bm p^2\over (\bm k-\bm p)^2}\right)\over -i\Omega+(\bar\eta_\perp\Lambda^{z-2}\bm k_\perp^2+\bar\eta_\parallel\Lambda^{z-2}\bm k_\parallel^2)+\lambda_\parallel\Lambda^{z-4}(\bm k_\parallel-\bm p_\parallel)^2(\bm k-\bm p)^2}\,,\nonumber\\
\eear
and the sum of the two becomes 
\be
+ig^2\Lambda^{2z-6+\epsilon}\int_p \psi_a(-p)\psi_r(p){\bm p}_\parallel^2 \bm p^2\int_{\bm k}{\left(1-{(\bm k_\parallel\cdot\bm p_\parallel)^2\over \bm k_\parallel^2\bm p_\parallel^2}\right)\over\left[(\bar\eta_\perp\Lambda^{z-2}\bm k_\perp^2+\bar\eta_\parallel\Lambda^{z-2}\bm k_\parallel^2)+\lambda_\parallel\Lambda^{z-4}\bm k_\parallel^2\bm k^2\right] \bm k^2}\,,
\ee
where we take a limit of a small external momentum compared to the loop momentum: $p\ll k$.

Here comes an important step of approximation in the model H which is self-consistent.
We will see that $\bar\eta_{\perp,\parallel}$ grows large in the renormalization group running, while $\lambda_\parallel$ approaches a finite fixed point in the IR: after many steps of renormalization we have $\bar\eta_{\perp,\parallel}\gg \lambda_\parallel$. Since the loop momentum $\bm k$ in the shell is near the cutoff $\Lambda$, the term with $\lambda_\parallel$ in the denominator is then negligible compared to the terms with $\bar\eta_{\perp,\parallel}$. This gives finally the expression
\bear
&&+ig^2\Lambda^{z-4+\epsilon}\int_p \psi_a(-p)\psi_r(p){\bm p}_\parallel^2 \bm p^2\int_{\bm k}{\left(1-{(\bm k_\parallel\cdot\bm p_\parallel)^2\over \bm k_\parallel^2\bm p_\parallel^2}\right)\over\left[(\bar\eta_\perp\bm k_\perp^2+\bar\eta_\parallel\bm k_\parallel^2)\right] \bm k^2}\,.\label{lambda}
\eear

We now describe the $\bm k$ integral in the above. We have
\be
\int_{\bm k}={1\over (2\pi)^{4-\epsilon}}\int_{\Lambda/b}^\Lambda d|\bm k| |\bm k|^{3-\epsilon} \int_{S^{3-\epsilon}} d\Omega\,,
\ee
and the integrand in (\ref{lambda}) is of a form
\be
{1\over |\bm k|^4} F(\Omega)\,,
\ee
with an angular function on the sphere $F(\Omega)$. The $|\bm k|$ integral gives
\be
\int_{\Lambda/b}^\Lambda {d|\bm k|\over |\bm k|} |\bm k|^{-\epsilon} =\Lambda^{-\epsilon}\log b+{\cal O}(\epsilon)\,,
\ee
and in the leading order in $\epsilon$-expansion, the angular integral can be performed in $\epsilon=0$ limit, that is, on the $S^3$ sphere. We parameterize the $\bm k$ space in this limit as
\bear
\bm k_1&=&|\bm k|\cos\theta\,,\nonumber\\
\bm k_2&=&|\bm k|\sin\theta\cos\phi\,,\nonumber\\
\bm k_3&=&|\bm k|\sin\theta\sin\phi\cos\chi\,,\nonumber\\
\bm k_4&=&|\bm k|\sin\theta\sin\phi\sin\chi\,,\label{s3}
\eear
with $0<(\theta,\phi)<\pi$ and $0<\chi<2\pi$, and the measure is
\be
d\Omega =\sin^2\theta\sin\phi d\theta d\phi d\chi\,.
\ee
The result of the angular integral depends on the dimension of the $\bm x_T$ space, that is, $d_T=1$ or $2$. For $d_T=1$, let's take $\bm k_\perp=\bm k_1$ and $\bm k_\parallel=(\bm k_2,\bm k_3,\bm k_4)$, and let $\bm p_\parallel$ point to $\bm k_4$ direction. The $F(\Omega)$ in this case is
\be
F(\Omega,\bar\eta_{\perp},\bar\eta_{\parallel})={(1-\sin^2\phi\sin^2\chi)\over \bar\eta_{\perp}\cos^2\theta +\bar\eta_\parallel\sin^2\theta}\,,\quad (d_T=1)\,.\label{f1}
\ee
The result should not depend on these choices as one can check.
For $d_T=2$, we choose $\bm k_\perp=(\bm k_1,\bm k_2)$ and $\bm k_\parallel=(\bm k_3,\bm k_4)$,
and $\bm p_\parallel$ pointing to the $\bm k_4$ direction. Then $F(\Omega)$ becomes
\be
F(\Omega,\bar\eta_\perp,\bar\eta_\parallel)={\cos^2\chi\over \bar\eta_\perp(\cos^2\theta+\sin^2\theta\cos^2\phi)+\bar\eta_\parallel\sin^2\theta\sin^2\phi}\,,\quad (d_T=2)\,.\label{f2}
\ee

With these the (\ref{lambda}) becomes at leading order in $\epsilon$
\bear
&&+i{g^2\Lambda^{z-4}\over (2\pi)^{4}}\int d\Omega \,F(\Omega,\bar\eta_\perp,\bar\eta_\parallel)\log b\int_p \psi_a(-p)\psi_r(p){\bm p}_\parallel^2 \bm p^2\nonumber\\
&=&{g^2\Lambda^{z-4}\over (2\pi)^{4}}\int d\Omega \,F(\Omega,\bar\eta_\perp,\bar\eta_\parallel)\log b\int dt\int d^{4-\epsilon}\bm x\,\,\psi_a(x)(-i\nabla_\parallel^2)(-\nabla^2)\psi_r(x)\,,
\eear
where the last line is a space-time expression. Looking at the diffusion term in the action (\ref{action}), this corresponds to a non-linear contribution we are looking for
\be
\delta \lambda_\parallel={g^2\over (2\pi)^{4}}\int d\Omega \,F(\Omega,\bar\eta_\perp,\bar\eta_\parallel)\log b\,.
\ee

We next compute the diagram in Figure \ref{fig2} for the renormalization of shear viscosities.
We have (neglecting an external frequency)
\bear 
g^2\Lambda^{2z-6+\epsilon}\int_p \bm j^i_{\parallel a}(-p)\bm j^j_{\parallel r}(p)\int_k {i(\bm k_\parallel^i-\bm p_\parallel^i)i(\bm k_\parallel^j(\bm p-\bm k)^2+(\bm p_\parallel^j-\bm k_\parallel^j)\bm k^2)2\lambda_\parallel\Lambda^{z-4}(\bm p_\parallel-\bm k_\parallel)^2\over
(-\omega-i\lambda_\parallel\Lambda^{z-4}\bm k_\parallel^2\bm k^2)(\omega^2+(\lambda_\parallel\Lambda^{z-4}(\bm p_\parallel-\bm k_\parallel)^2(\bm p-\bm k)^2)^2)}\,,\nonumber\\
\eear
and performing the $\omega$-integration, we obtain
\be
-i{g^2\over\lambda_\parallel}\Lambda^{z-2+\epsilon}\int_p \bm j^i_{\parallel a}(-p)\bm j^j_{\parallel r}(p)\int_{\bm k}
{\bm k_\parallel^i \bm k_\parallel^j\left(1-{\bm k^2\over (\bm k-\bm p)^2}\right)\over (\bm k_\parallel-\bm p_\parallel)^2(\bm k-\bm p)^2+\bm k_\parallel^2\bm k^2}\,.\label{shear1}
\ee
Recall that $\bm j_\parallel(p)$ lies in $\bm k_\parallel$ space, {\it and} is also perpendicular to $\bm p_\parallel$ (and hence  to $\bm p$) inside the $\bm k_\parallel$ space to be a shear component.
Therefore we can replace in the numerator
\be
\bm k_\parallel^i \bm k_\parallel^j\to {1\over 3-d_T-\epsilon} \delta_{\parallel\perp}^{ij} \bm k^2_{\parallel\perp} \,,
\ee
where $\bm k_{\parallel\perp}$ is the subspace inside $\bm k_\parallel$ perpendicular to $\bm p_\parallel$, and its dimension is $d_L-1=3-d_T-\epsilon$. Also expanding the remaining part of the integrand up to the first relevant leading order, 
\be
{\left(1-{\bm k^2\over (\bm k-\bm p)^2}\right)\over (\bm k_\parallel-\bm p_\parallel)^2(\bm k-\bm p)^2+\bm k_\parallel^2\bm k^2}\approx {1\over 2\bm k_\parallel^2\bm k^4}\left(\bm p^2-2 {(\bm k_\parallel\cdot\bm p_\parallel)^2\over \bm k_\parallel^2}-6 {(\bm k\cdot\bm p)^2\over \bm k^2}\right)\,,
\ee
the generated action (\ref{shear1}) then becomes
\be
-i{g^2\over \lambda_\parallel}\Lambda^{z-2+\epsilon}{1\over 2(3-d_T-\epsilon)}\int_p \bm j_{\parallel a}(-p)\cdot\bm j_{\parallel r}(p)\int_{\bm k}{\bm k_{\parallel\perp}^2\over \bm k_\parallel^2\bm k^4} \left(\bm p^2-2 {(\bm k_\parallel\cdot\bm p_\parallel)^2\over \bm k_\parallel^2}-6 {(\bm k\cdot\bm p)^2\over \bm k^2}\right)\,,
\ee
which is logarithmically sensitive to the cutoff. The $\bm k$ integral is done similarly as before. There are two types of terms in the result: a term proportional to $\bm p_\perp^2$ and the other proportional to $\bm p_\parallel^2$. They correspond to a renormalization of $\bar\eta_\perp$ and $\bar\eta_\parallel$ respectively. The first type of term is
\be
-i{g^2\over \lambda_\parallel}\Lambda^{z-2}{1\over 2(3-d_T-\epsilon)}{1\over (2\pi)^{4-\epsilon}}\int d\Omega\,\,{\bm k_{\parallel\perp}^2\over \bm k_\parallel^2} \left(1-6 {(\bm k_\perp\cdot\bm p_\perp)^2\over \bm k^2\bm p_\perp^2}\right)\log b \int_p \bm j_{\parallel a}(-p)\cdot\bm j_{\parallel r}(p)\bm p_\perp^2\,,
\ee
and the second type is
\be
-i{g^2\Lambda^{z-2}\over \lambda_\parallel}{1\over 2(3-d_T-\epsilon)}{1\over (2\pi)^{4-\epsilon}}\int d\Omega {\bm k_{\parallel\perp}^2\over \bm k_\parallel^2} \left(1-2{(\bm k_\parallel\cdot\bm p_\parallel)^2\over \bm k_\parallel^2\bm p_\parallel^2}-6 {(\bm k_\parallel\cdot\bm p_\parallel)^2\over \bm k^2\bm p_\parallel^2}\right) \log b\int_p \bm j_{\parallel a}(-p)\cdot\bm j_{\parallel r}(p)\bm p_\parallel^2\,.
\ee
What remains in these terms is an angular integration on $S^3$ in $\epsilon=0$ limit. Using the parametrization (\ref{s3}), 
we obtain for $d_T=1$ case
\bear
{1\over (2\pi)^{4}}\int d\Omega\,\,{\bm k_{\parallel\perp}^2\over \bm k_\parallel^2} \left(1-6 {(\bm k_\perp\cdot\bm p_\perp)^2\over \bm k^2\bm p_\perp^2}\right)&=&-{1\over 8\pi^2} {1\over 3}\,,\nonumber\\
 {1\over (2\pi)^4}\int d\Omega {\bm k_{\parallel\perp}^2\over \bm k_\parallel^2} \left(1-2{(\bm k_\parallel\cdot\bm p_\parallel)^2\over \bm k_\parallel^2\bm p_\parallel^2}-6 {(\bm k_\parallel\cdot\bm p_\parallel)^2\over \bm k^2\bm p_\parallel^2}\right)&=&-{1\over 8\pi^2} {1\over 5} \,, \eear
and for $d_T=2$ case,
\bear
{1\over (2\pi)^{4}}\int d\Omega\,\,{\bm k_{\parallel\perp}^2\over \bm k_\parallel^2} \left(1-6 {(\bm k_\perp\cdot\bm p_\perp)^2\over \bm k^2\bm p_\perp^2}\right)&=& -{1\over 8\pi^2} {1\over 4} \,,\nonumber\\
 {1\over (2\pi)^4}\int d\Omega {\bm k_{\parallel\perp}^2\over \bm k_\parallel^2} \left(1-2{(\bm k_\parallel\cdot\bm p_\parallel)^2\over \bm k_\parallel^2\bm p_\parallel^2}-6 {(\bm k_\parallel\cdot\bm p_\parallel)^2\over \bm k^2\bm p_\parallel^2}\right)&=& -{1\over 8\pi^2} {1\over 8}\,.
 \eear
From these, we finally obtain the non-linear contributions to the renormalization of $\bar\eta_{\perp}$ and $\bar\eta_\parallel$,
\bear
\delta\bar\eta_\perp={1\over 12}{1\over 8\pi^2}{g^2\over\lambda_\parallel}\log b\,,\quad \delta\bar\eta_\parallel={1\over 20}{1\over 8\pi^2}{g^2\over\lambda_\parallel}\log b\,,&& (d_T=1)\\
\delta\bar\eta_\perp={1\over 8}{1\over 8\pi^2}{g^2\over\lambda_\parallel}\log b\,,\quad \delta\bar\eta_\parallel={1\over 16}{1\over 8\pi^2}{g^2\over\lambda_\parallel}\log b\,.&& (d_T=2)
\eear

\begin{figure}[t]
	\centering
	\includegraphics[width=10cm]{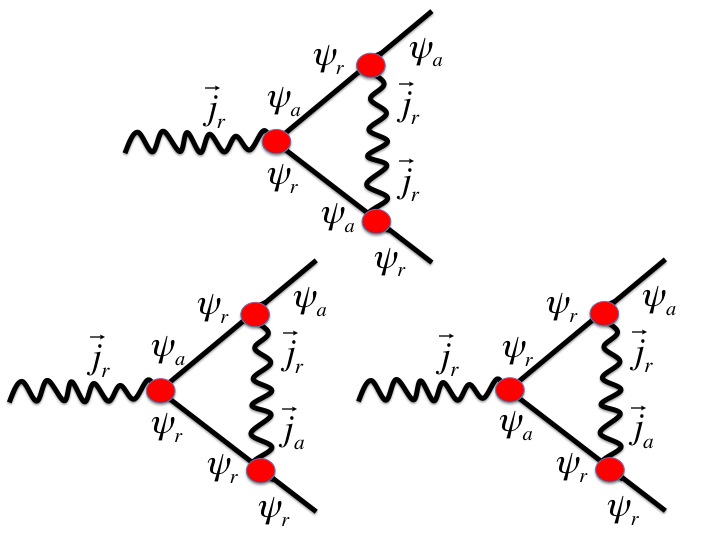}
		\caption{The real-time diagrams that are relevant for $\delta g$. It turns out that there is no correction to $\delta g$ from these diagrams.\label{fig3}}
\end{figure}
The non-linear contributions to $\delta g$ can potentially arise from the diagrams in Figure \ref{fig3}.
Explicit computations show that these diagrams generate only higher dimensional terms, and a correction to $\delta g$ is absent \cite{Siggia:1976zz}.
 
In summary of all these, the renormalization group equations are
\bear
{d\lambda_\parallel\over d\log b}&=& (z-4+\eta)\lambda_\parallel+{g^2\over (2\pi)^4}\int d\Omega \,\,F(\Omega,\bar\eta_\perp,\bar\eta_\parallel)\,,\nonumber\\
{d\bar\eta_\perp\over d\log b}&=&(z-2)\bar\eta_\perp+{1\over 4(4-d_T)}{1\over 8\pi^2}{g^2\over\lambda_\parallel}\,,\nonumber\\
{d\bar\eta_\parallel\over d\log b}&=&(z-2)\bar\eta_\parallel+{1\over 4(6-d_T)}{1\over 8\pi^2}{g^2\over\lambda_\parallel}\,,\nonumber\\
{dg\over d\log b}&=& (z-3+\epsilon/2)g\,,\label{flow}
\eear
 where the angular function $F(\Omega,\bar\eta_\perp,\bar\eta_\parallel)$ is given in (\ref{f1}) and (\ref{f2}).

\section{Fixed point and the critical exponents \label{sec4}}

Following the model H analysis \cite{Siggia:1976zz,Hohenberg:1977ym}, we define the two coupling constants,
\be
f_\perp\equiv {1\over 8\pi^2}{g^2\over\bar\eta_\perp\lambda_\parallel}\,,\quad 
f_\parallel\equiv {1\over 8\pi^2}{g^2\over\bar\eta_\parallel\lambda_\parallel}\,.
\ee
Using these variables, the flow equations (\ref{flow}) can be written as
\bear
\lambda_\parallel'&=&b^{z-4+\eta}\lambda_\parallel\left(1+{1\over 2\pi^2}\int d\Omega \,\,F\left(\Omega,1/f_\perp,1/f_\parallel\right)\log b\right)\approx \lambda_\parallel b^{z-4+\eta+{1\over 2\pi^2}\int d\Omega \,\,F\left(\Omega,1/f_\perp,1/f_\parallel\right)}\,,\nonumber\\
\bar\eta_\perp'&=& b^{z-2}\bar\eta_\perp\left(1+{1\over 4(4-d_T)}f_\perp\log b\right)\approx \bar\eta_\perp b^{z-2+{1\over 4(4-d_T)}f_\perp}\,,\nonumber\\
\bar\eta_\parallel'&=& b^{z-2}\bar\eta_\parallel\left(1+{1\over 4(6-d_T)}f_\parallel\log b\right)\approx \bar\eta_\parallel b^{z-2+{1\over 4(6-d_T)}f_\parallel}\,,\nonumber\\
g'&=& b^{z-3+\epsilon/2} g\,,
\eear
where the primed parameters are the renormalized ones after performing one step of integrating a momentum shell and rescaling, and the front factors of $b$ are from naive dimensional scaling, which should be undone for the physical parameters measured in the original coordinate-time. 
Therefore the physical parameters receive contributions only from the non-linear effects.
We see that these effects are given solely by the parameters $f_\perp$ and $f_\parallel$.
More explicitly, for infinitesimal $\log b$,
\bear
\lambda_\parallel^{phys}&=&\lambda_\parallel^0 \exp\left[{1\over 2\pi^2}\int_0^{\log(\Lambda\xi)}d\log b\,\int d\Omega \,\,F\left(\Omega,1/f_\perp(\log b),1/f_\parallel(\log b)\right)\right]\,,\nonumber\\
\bar\eta_\perp^{phys}&=&\bar\eta_\perp^0\exp\left[{1\over 4(4-d_T)}\int_0^{\log(\Lambda\xi)}d\log b\,\,f_\perp(\log b)\right]\,,\nonumber\\
\bar\eta_\parallel^{phys}&=&\bar\eta_\parallel^0\exp\left[{1\over 4(6-d_T)}\int_0^{\log(\Lambda\xi)}d\log b\,\,f_\parallel(\log b)\right]\,,\label{physical}
\eear
where $\lambda^0_\parallel$, $\bar\eta_{\perp,\parallel}^0$ are the parameters at the cutoff $\Lambda$ in the original theory, and $f_{\perp,\parallel}(\log b)$ are the running couplings by solving the flow equations (\ref{flow}). If there exists an attractive fixed point for $f_{\perp,\parallel}$ as we will see shortly, the integral is dominated by the fixed point value for a large $\Lambda\xi\gg 1$ in a scaling regime. This motivates us to look at the flow equations for $f_{\perp,\parallel}$.
From (\ref{flow}), we obtain
\bear
{d f_\perp\over d\log b}&=&(\epsilon-\eta)f_\perp-{1\over 4(4-d_T)}f_\perp^2-f_\perp{1\over 2\pi^2}\int d\Omega\,\,F(\Omega,1/f_\perp,1/f_\parallel)\,,\nonumber\\
{d f_\parallel\over d\log b}&=&(\epsilon-\eta)f_\perp-{1\over 4(6-d_T)}f_\parallel^2-f_\parallel{1\over 2\pi^2}\int d\Omega\,\,F(\Omega,1/f_\perp,1/f_\parallel)\,.
\eear
It is not difficult to find a fixed point $(f_\perp^*,f_\parallel^*)$ that makes the right-hand side of the above equations vanish. To leading order in $\epsilon$, let's ignore $\eta\sim \epsilon^2$.
First define a constant $C$ by
\be
C\epsilon={1\over 2\pi^2}\int d\Omega\,\,F(\Omega,1/f^*_\perp,1/f^*_\parallel)\,,\label{con}
\ee
then the fixed point values are
\be
f_\perp^*=4(4-d_T)(1-C)\epsilon\,,\quad f_\parallel^*=4(6-d_T)(1-C)\epsilon\,,
\ee
and inserting these back to (\ref{con}) gives a self-consistent equation for $C$,
\be
C={1\over 2\pi^2}\int d\Omega\,\,F\left(\Omega,{1\over 4(4-d_T)},{1\over 4(6-d_T)}\right)(1-C)\,,
\ee
that is,
\be
C={{1\over 2\pi^2}\int d\Omega\,\,F\left(\Omega,{1\over 4(4-d_T)},{1\over 4(6-d_T)}\right)\over 1+{1\over 2\pi^2}\int d\Omega\,\,F\left(\Omega,{1\over 4(4-d_T)},{1\over 4(6-d_T)}\right)}\,.
\ee
With the expression for $F(\Omega,x,y)$ in (\ref{f1}) and (\ref{f2}), we obtain
\be
C\approx 0.921\quad (d_T=1)\,,\qquad C\approx 0.847\quad (d_T=2)\,.
\ee

Once we find the fixed point, the physical parameters are obtained by integrating (\ref{physical}) as
\bear
\lambda_\parallel^{phys}&=&\lambda_\parallel^0 (\Lambda\xi)^{C\epsilon}\,,\nonumber\\
\bar\eta_\perp^{phys}&=&\bar\eta_\perp^0 (\Lambda\xi)^{f_\perp^*/4(4-d_T)}=\bar\eta_\perp^0 (\Lambda\xi)^{(1-C)\epsilon}\,,\nonumber\\
\bar\eta_\parallel^{phys}&=&\bar\eta_\parallel^0 (\Lambda\xi)^{f_\parallel^*/4(6-d_T)}=\bar\eta_\parallel^0 (\Lambda\xi)^{(1-C)\epsilon}\,.
\eear
We find that $\bar\eta_\perp$ and $\bar\eta_\parallel$ share the same critical behavior, and this justifies why we need to keep both.
The critical exponents $x_\lambda$ and $x_{\bar\eta}$ are defined by
\be
\lambda_\parallel^{phys}\sim \xi^{x_\lambda}\,,\quad \bar\eta_{\perp,\parallel}\sim \xi^{x_{\bar\eta}}\,,
\ee
and we have up to order $\epsilon$,
\be
x_\lambda=C\epsilon\,,\quad x_{\bar\eta}=(1-C)\epsilon\,,
\ee
which satisfies the well-known scaling relation in the model H; $x_\lambda+x_{\bar\eta}=\epsilon-\eta$ \cite{Siggia:1976zz,Hohenberg:1977ym}.
For a comparison, the original model H has the value $C=18/19\approx 0.947$.
We find that $d_T=1$ case is somewhat close to the original model H, but for $d_T=2$ case the difference in $C$ is about $10\%$ which is significant. We recall that $d_T=2$ case is relevant for the QCD critical point in a background magnetic field. It is also easy to see that $\bar\eta_{\perp,\parallel}$
grows much faster than $\lambda_\parallel$, which justifies the approximation in obtaining the equation (\ref{lambda}).

The relaxation frequency for the charge diffusion mode in hydro-regime $k\ll \xi^{-1}$ is given by
\be
\omega\sim{1\over\chi}{\lambda_\parallel^{phys}}k_\parallel^2\,,
\ee
where $\chi\sim\xi^{2-\eta}$ is the susceptibility that is the inverse of the parameter $r^{phys}$.
Matching to the critical regime near $k\sim \xi^{-1}$, we have up to order $\epsilon$
\be
\omega\sim k^{4-x_\lambda}\,,
\ee
as the relaxation frequency in the scaling regime $k\gg \xi^{-1}$.
For the shear modes, we instead have
\be
\omega\sim \bar\eta k^2\sim k^{2-x_{\bar\eta}}\,.
\ee
Since the charge diffusion modes relax more slowly than the shear modes, they define the critical slowing down. This gives the dynamic critical exponent \cite{Siggia:1976zz,Hohenberg:1977ym}
\be
z=4-x_\lambda=4-C\epsilon\,.
\ee
With this, the flow equation for $\lambda_\parallel$ in (\ref{flow}) has a finite fixed point for $\lambda_\parallel^*<\infty$. For $d_T=2$ case, we have $z\approx 3.15$ when $\epsilon=1$ (three dimensions).
This is notably larger than the original model H value, $z\approx 3.05$.

As a further direction, one could try to compute other refined quantities in these models such as scaling functions. It should also be possible to go to a next order in $\epsilon$ expansion using the method in Ref.\cite{Wilson:1971vs}


\vskip 1cm \centerline{\large \bf Acknowledgment} \vskip 0.5cm
We thank Rob Pisarski, Misha Stephanov and Ismail Zahed for discussions.
This work is partially supported by the U.S. Department of Energy, Office of Science, Office of Nuclear Physics, within the framework of the Beam Energy Scan Theory (BEST) Topical Collaboration.

 \vfil


\begin{thebibliography}{99} \frenchspacing

\bibitem{Siggia:1976zz} 
  E.~D.~Siggia, B.~I.~Halperin and P.~C.~Hohenberg,
  ``Renormalization-group treatment of the critical dynamics of the binary-fluid and gas-liquid transitions,''
  Phys.\ Rev.\ B {\bf 13}, 2110 (1976).

\bibitem{Hohenberg:1977ym} 
  P.~C.~Hohenberg and B.~I.~Halperin,
  ``Theory of Dynamic Critical Phenomena,''
  Rev.\ Mod.\ Phys.\  {\bf 49}, 435 (1977).
  
\bibitem{Wilson:1971dc} 
  K.~G.~Wilson and M.~E.~Fisher,
  ``Critical exponents in 3.99 dimensions,''
  Phys.\ Rev.\ Lett.\  {\bf 28}, 240 (1972).
  
\bibitem{Wilson:1973jj} 
  K.~G.~Wilson and J.~B.~Kogut,
  ``The Renormalization group and the epsilon expansion,''
  Phys.\ Rept.\  {\bf 12}, 75 (1974).
  
\bibitem{Kadanoff:1968zz} 
  L.~P.~Kadanoff and J.~Swift,
  ``Transport Coefficients near the Liquid-Gas Critical Point,''
  Phys.\ Rev.\  {\bf 166}, 89 (1968).
 
  \bibitem{kawasaki}
  K. Kawasaki, Ann. Phys. (N.Y.) 61, 1 (1970).
  
\bibitem{Li:2017tgi} 
  S.~Li and H.~U.~Yee,
  ``Shear Viscosity of Quark-Gluon Plasma in Weak Magnetic Field in Perturbative QCD: Leading Log,''
  arXiv:1707.00795 [hep-ph].
  
\bibitem{Stephanov:1999zu} 
  M.~A.~Stephanov, K.~Rajagopal and E.~V.~Shuryak,
  ``Event-by-event fluctuations in heavy ion collisions and the QCD critical point,''
  Phys.\ Rev.\ D {\bf 60}, 114028 (1999).
  
\bibitem{Berdnikov:1999ph} 
  B.~Berdnikov and K.~Rajagopal,
  ``Slowing out-of-equilibrium near the QCD critical point,''
  Phys.\ Rev.\ D {\bf 61}, 105017 (2000).


\bibitem{Son:2004iv} 
  D.~T.~Son and M.~A.~Stephanov,
  ``Dynamic universality class of the QCD critical point,''
  Phys.\ Rev.\ D {\bf 70}, 056001 (2004).
  
\bibitem{Wilson:1971vs} 
  K.~G.~Wilson,
  ``Feynman graph expansion for critical exponents,''
  Phys.\ Rev.\ Lett.\  {\bf 28}, 548 (1972).
  
  
  
\end{thebibliography}
\end{document}